\documentclass{moriond}

\usepackage{amsmath,amssymb}

\def\Journal#1#2#3#4{{#1} {\bf #2}, #3 (#4)}

\def\be{\begin{equation}}
\def\ee{\end{equation}}
\def\bea{\begin{eqnarray}}
\def\eea{\end{eqnarray}}

\newcommand{\Photo}{\includegraphics[height=35mm]{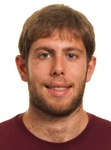}}

\begin{document}

\title{Impact of $\mathbf{A_{0}}$ data on the Higgs boson production cross section at the LHC}

\author{ Francesco Giuli }

\address{University of Rome ``Tor Vergata'' and INFN Section of Rome 2,\\ Via della Ricerca Scientifica 1, 00133, Rome, Italy}

\maketitle\abstracts{
In this talk, we present a way to improve the accuracy of theoretical predictions for Higgs boson production cross sections at the LHC using the measurements of lepton angular distributions. In this regards, we exploit the sensitivity of the lepton angular coefficient associated with the longitudinal $Z$-boson polarization to the parton density function (PDF) for gluons resolved from the incoming protons, in order to constrain the Higgs boson cross section from gluon fusion processes. We find that high-statistics determinations of the longitudinally polarized angular coefficient at the LHC Run 3 and high-luminosity HL-LHC improve the PDF systematics of the Higgs boson cross section predictions by 50~\% over a broad range of Higgs boson rapidities. This study has been conducted using the open-source fitting framework \texttt{xFitter}. This talk refers to the following paper~\cite{Amoroso:2020fjw}.
}

\section{Introduction}
One of the key studies of the current and forthcoming physics 
programs at the Large Hadron Collider (LHC) is to precisely measure the Higgs sector of the Standard Model (SM). At the LHC, the dominant mechanism for the production of Higgs bosons is through gluon-fusion. In this work, we investigate the sensitivity to the gluon PDF via ${\cal O} ( \alpha_s)$ contributions. In particular, we consider Drell-Yan (DY) charged lepton-pair production via $Z/\gamma^{*}$ exchange and the focus of our study is the $A_{0}$ angular coefficient, defined as the ratio of the longitudinal electroweak boson cross section to the unpolarized cross section:
\begin{equation} 
\label{A0def} 
A_0 ( s , M, Y, p_T) = { {2 d \sigma^{(L)} / dM dY dp_T } \over   { d \sigma / dM dY dp_T} } . 
\end{equation}
It has been shown that this coefficient is perturbatively stable~\cite{Chang:2017kuv,Gauld:2017tww} and it has been presicely measured at Tevatron~\cite{Aaltonen:2011nr} and the LHC~\cite{Khachatryan:2015paa,Aad:2016izn}, as well as at fixed-target experiments~\cite{Guanziroli:1987rp,Conway:1989fs,Zhu:2006gx,Zhu:2008sj}.\\
In order to study the impact of high-precision $A_0$ measurements on the 
Higgs boson production cross section and to evaluate the reduction in PDF uncertainties, we implement the NLO {\tt{MadGraph5\_aMC@NLO}}~\cite{Alwall:2014hca} $A_0$ calculation into the open-source fit platform {\tt{xFitter}}~\cite{Alekhin:2014irh}.

\section{Fit to ATLAS 8 TeV data}
We perform NLO fits to the $\sqrt{s} = 8$ TeV ATLAS measurements~\cite{Aad:2016izn} for the angular coefficient $A_{0}$, using the unregularised data and including the covariance matrices of the experimental and PDF uncertainties. The considered PDF sets are: CT18nnlo, CT18Annlo~\cite{Hou:2019efy}, MSHT20nnlo~\cite{Bailey:2020ooq}, NNPDF3.1nnlo~\cite{Ball:2017nwa}, HERAPDF2.0nnlo~\cite{Abramowicz:2015mha} and ABMP16nnlo~\cite{Alekhin:2017kpj}. A very good description of data is achieved for all PDF sets, and the results for the $\chi^2$  values are reported in Tab.~\ref{chi2_8TeV}.
\begin{table}[!th]\small
\centering
\begin{tabular}{|c|c|}
\hline
PDF set & Total $\chi^2$/d.o.f.\\
\hline
CT18NNLO & 59/53 \\
\hline
CT18Annlo & 44/53 \\
\hline
MSHT20nnlo{\_}as118 & 59/53 \\
\hline
NNPDF31{\_}nnlo{\_}as{\_}0118{\_}hessian & 60/53 \\
\hline
ABMP16{\_}5{\_}nnlo & 62/53 \\
\hline
HERAPDF20{\_}NNLO{\_}EIG & 60/53\\
\hline
\end{tabular}
\caption{The $\chi^2$ values per degrees of freedom from NLO fits to $A_{0}$ data for different collinear PDF sets. PDF uncertainties are evaluated at the 68\% CL.}
\label{chi2_8TeV}
\end{table} 

\section{Gluon PDF profiling}
\begin{figure}[t!]
\begin{center}
\includegraphics[width=0.34\textwidth]{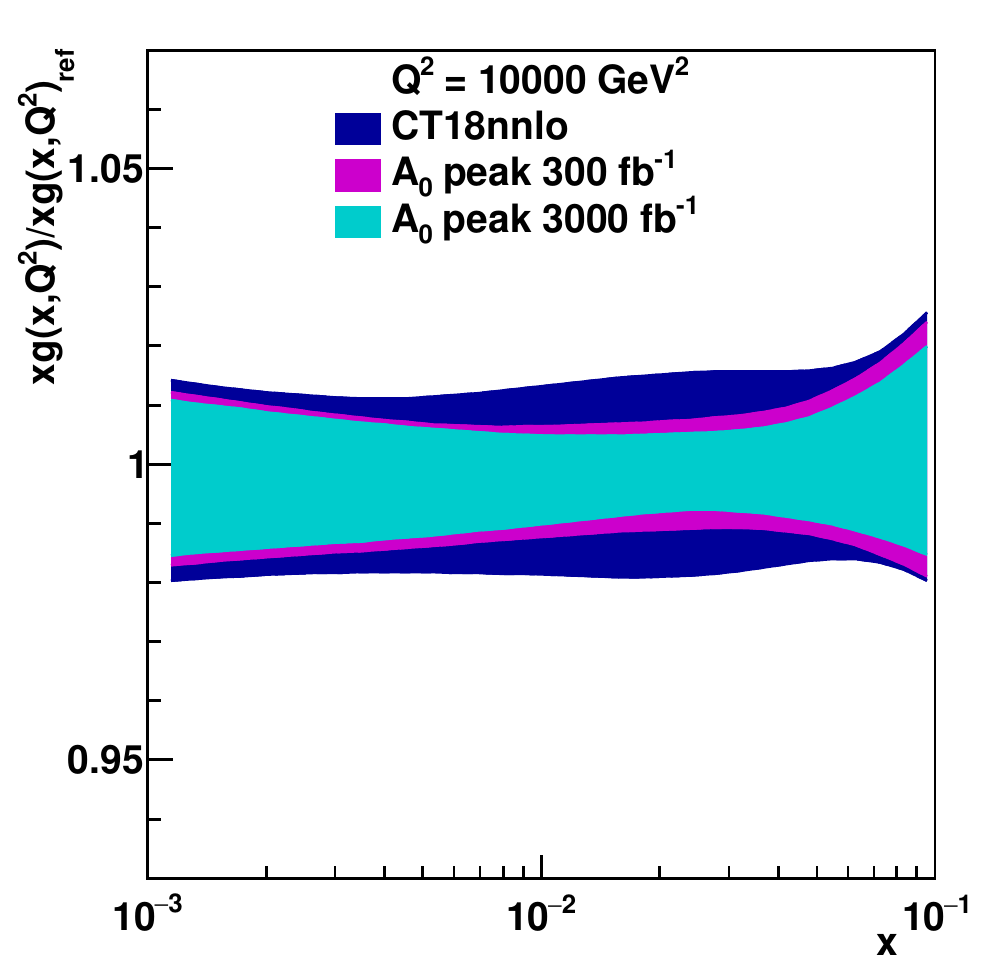}
\includegraphics[width=0.34\textwidth]{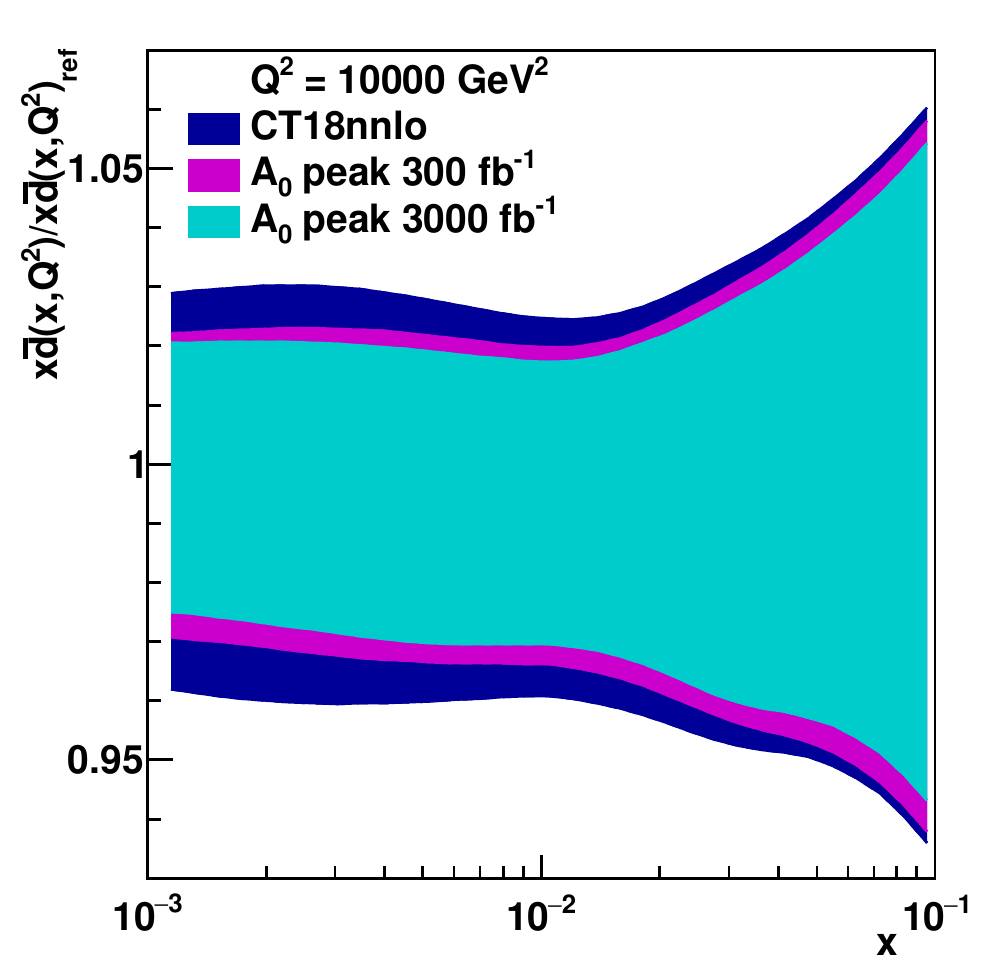}
\caption{Original CT18nnlo (dark blue) and profiled distributions using $A_0$ pseudodata corresponding to integrated luminosities of 300~fb$^{-1}$ (magenta) and 3~ab$^{-1}$ (cyan) for 80 $< M <$ 100~GeV and $|Y|<3.5$. Results for gluon ($xg$) and $d$-type ($x\bar{d}$) sea-quark densities are shown for $Q^{2}$ = 10$^{4}$~GeV$^{2}$. Bands represent PDF uncertainties, shown at the 68\% CL.}
\label{fig:CT18_13TeV_Second}
\end{center}
\end{figure}

\begin{figure}[t!]
\begin{center}
\includegraphics[width=0.34\textwidth]{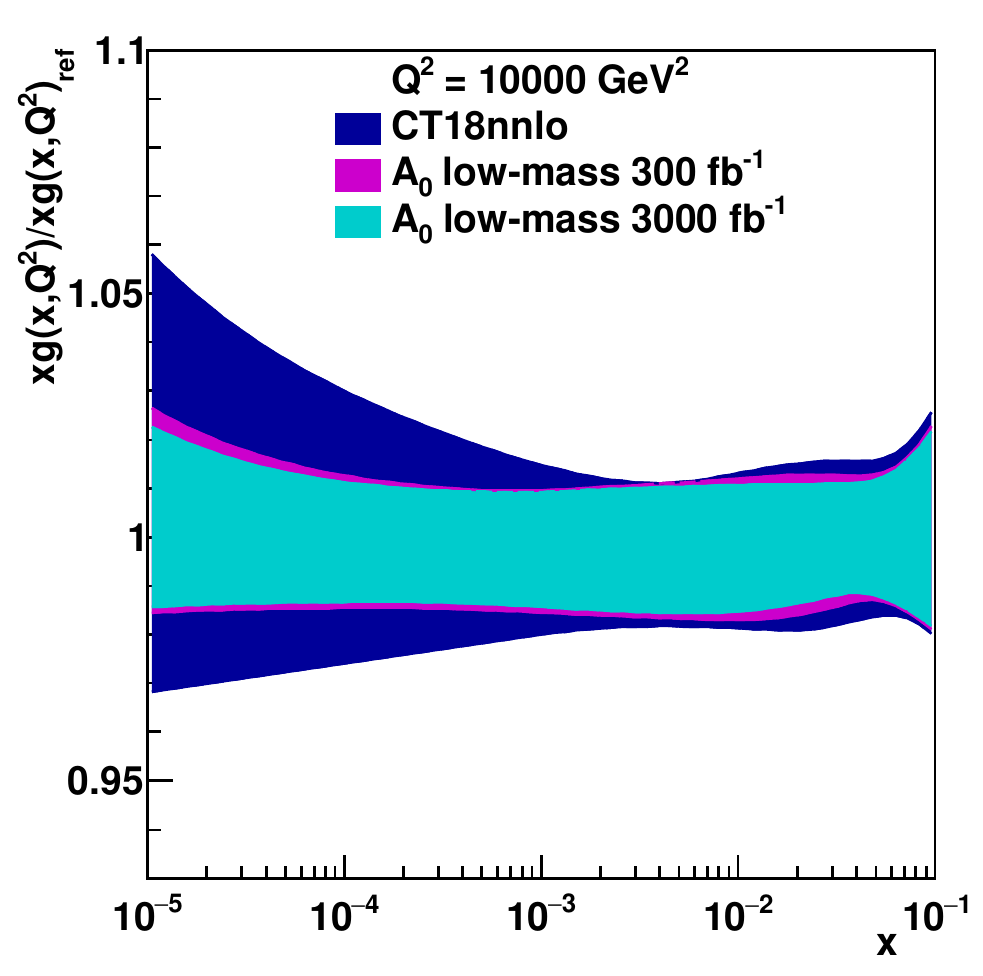}
\includegraphics[width=0.34\textwidth]{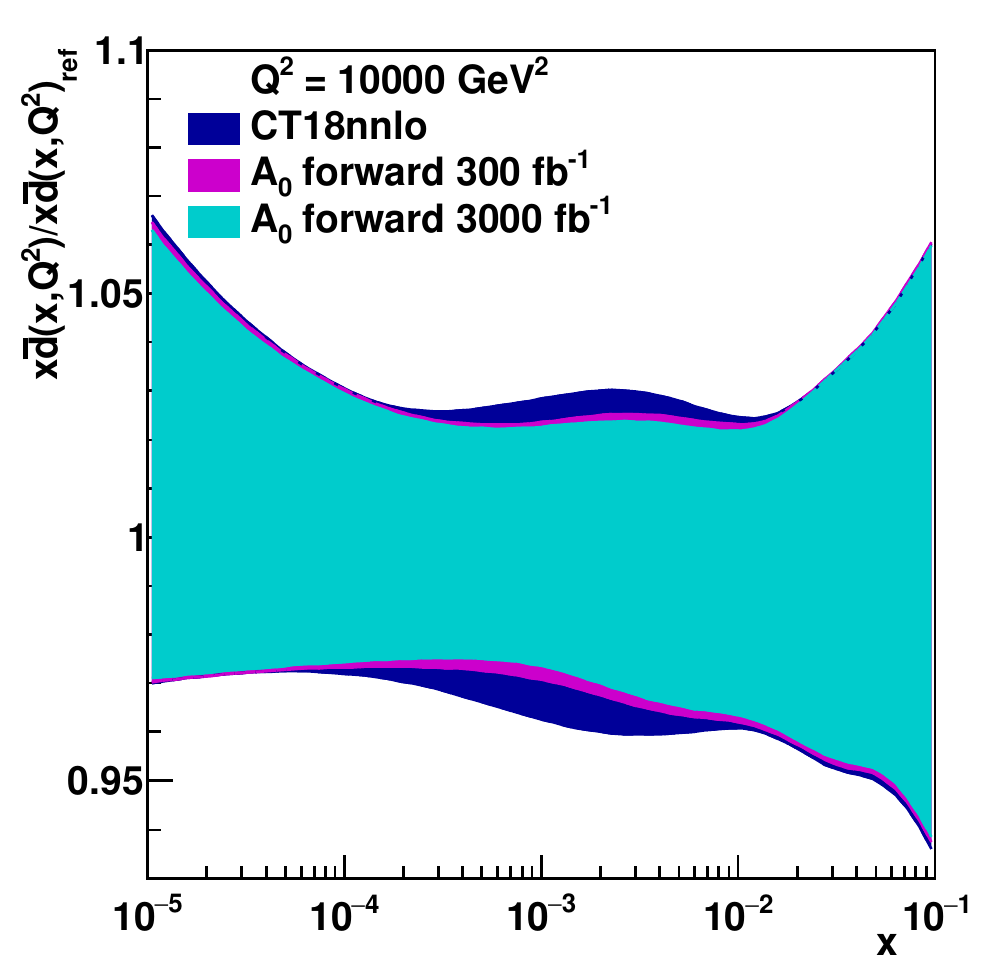}
\caption{Original CT18nnlo (dark blue) and profiled distributions using $A_0$ pseudodata corresponding to integrated luminosities of 300~fb$^{-1}$ (magenta) and 3~ab$^{-1}$ (cyan) for the low mass (left) and LHCb phase space (right). Results for gluon ($xg$) and $d$-type ($x\bar{d}$) sea-quark densities are shown for $Q^{2}$ = 10$^{4}$~GeV$^{2}$.  Bands represent PDF uncertainties, shown at the 68\% CL.}
\label{fig:CT18_13TeV_Second_appendix}
\end{center}
\end{figure}

We generate $A_0$ pseudodata at $\sqrt{s} = 13$~TeV for two projected luminosity scenarios of 300 fb$^{-1}$ (Run III scenario) and 3 ab$^{-1}$ (HL-LHC scenario), and apply  
the profiling technique~\cite{Paukkunen:2014zia,Camarda:2015zba} to evaluate the PDF uncertainties.\\
We perform the analysis in the  mass region  80 $< M <$ 100~GeV around the $Z$-boson peak and  rapidity region $|Y| <$ 3.5. The results are reported in Fig.~\ref{fig:CT18_13TeV_Second}. It is visible the largest reduction of uncertainties from the high-luminosity $A_0$ profiling occurs for the gluon density and for the $d$ sea-quark densities.\\ 
We also perform the profiling analysis in the low mass region  4 $< M <$ 8~GeV and forward rapidity region 2.0 $< |Y| <$ 4.5. We report the results in Fig.~\ref{fig:CT18_13TeV_Second_appendix}. The largest reduction of uncertainties observed in this case for the gluon distribution is at $x<0.001$. For the LHCb phase space, the most relevant improvements
are for the sea quark PDFs, e.g. $\bar{d}$ at $x\sim 0.001$.

\section{Impact on the Higgs cross section}
\begin{figure}[t!]
\begin{center}
\includegraphics[width=0.46\textwidth]{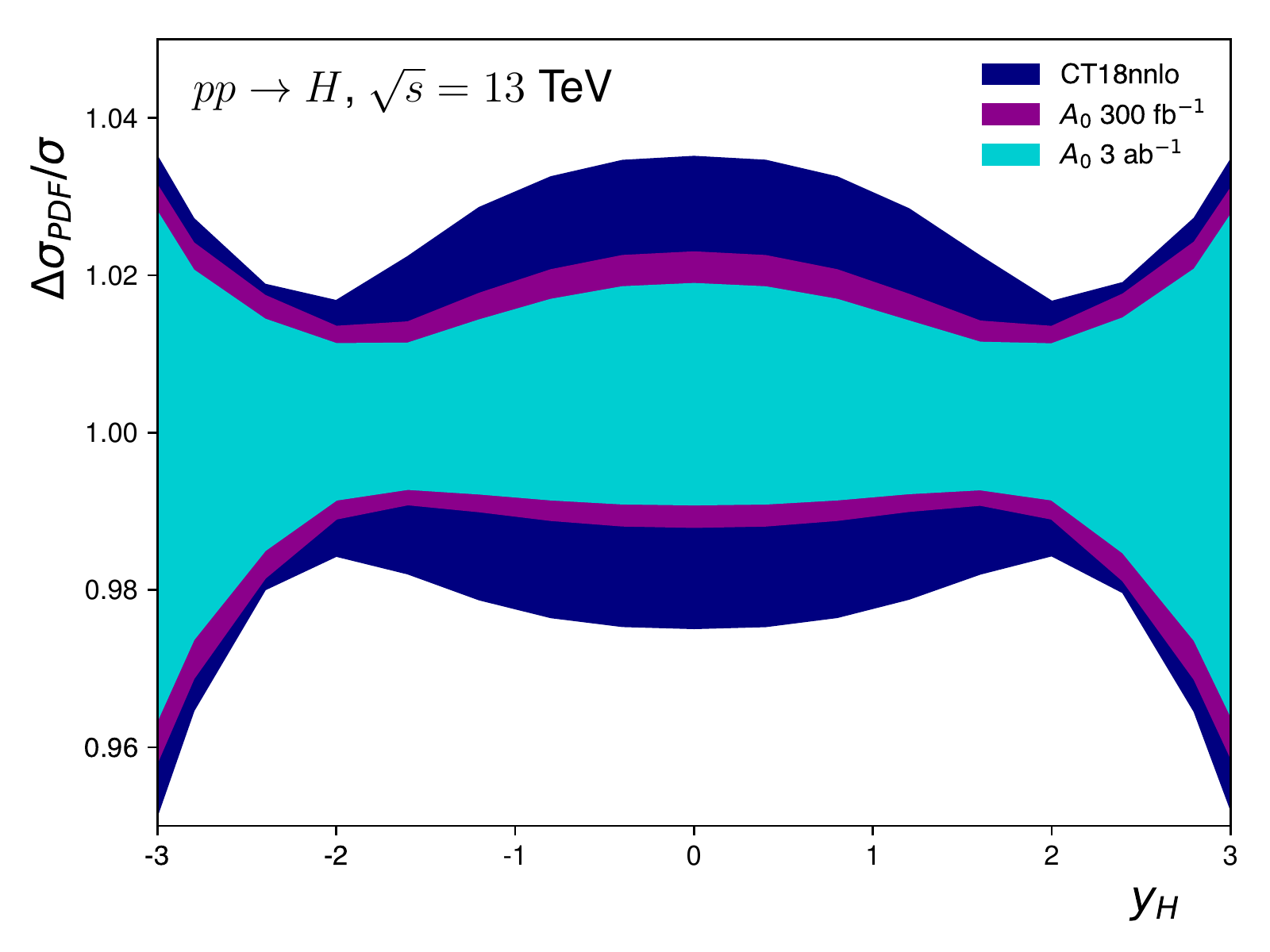}
\includegraphics[width=0.46\textwidth]{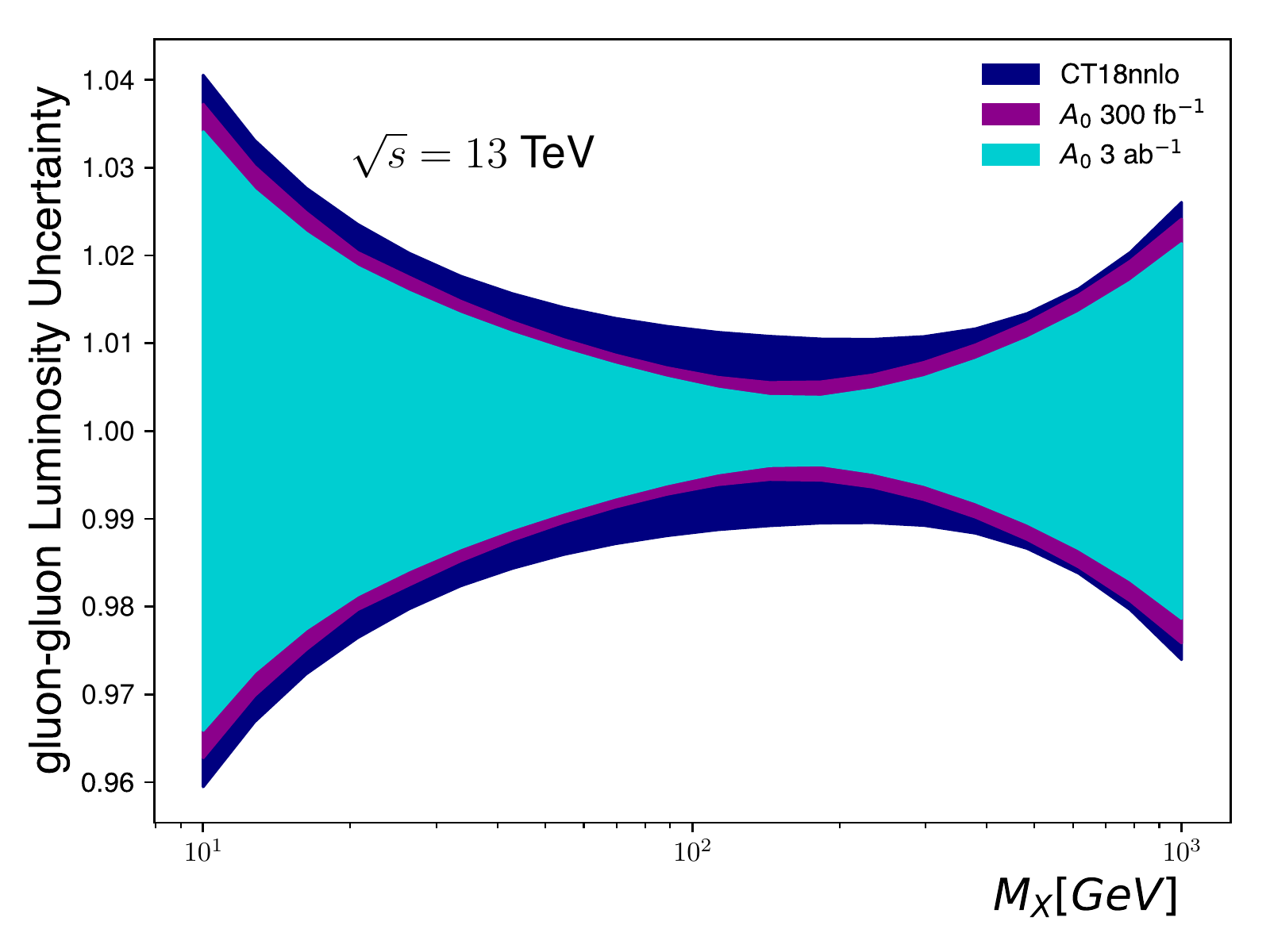}
\end{center}
\caption{Ratio of PDF uncertainties for the gluon-gluon fusion SM Higgs boson cross-section (left) and the gluon-gluon luminosity evaluated at $\sqrt{s}=$13~TeV (right). The dark blue band shows the uncertainties of the CT18nnlo PDF set, reduced to 68\% CL coverage. The magenta and 
cyan bands show the uncertainties of the CT18nnlo including constraints from the $A_0$ measurement and assuming 300~fb$^{-1}$ and 3~ab$^{-1}$, respectively.}
\label{fig:HiggsRapidity}
\end{figure}

\begin{figure}[t!]
\begin{center}
\includegraphics[width=0.46\textwidth]{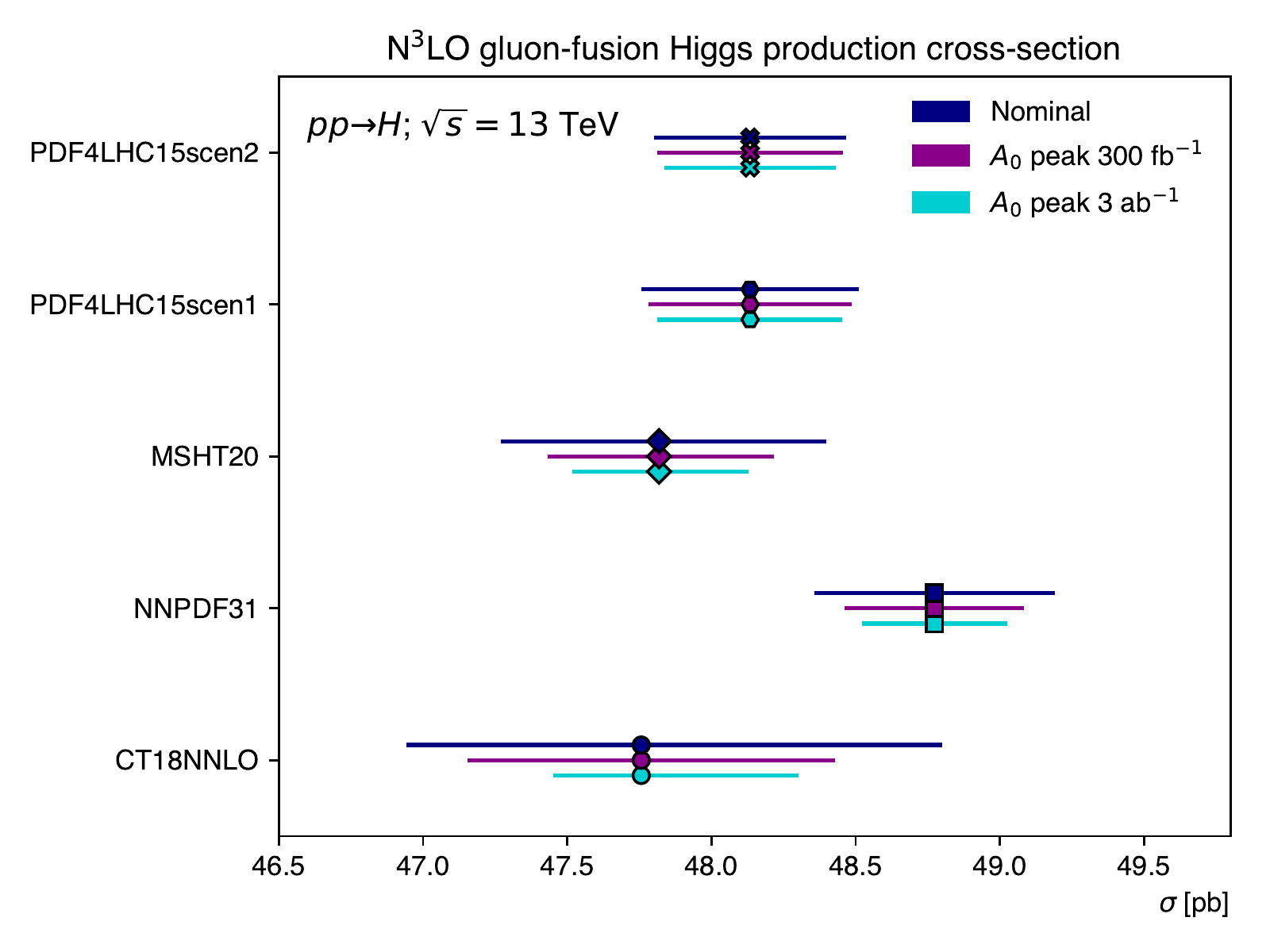}
\includegraphics[width=0.34\textwidth]{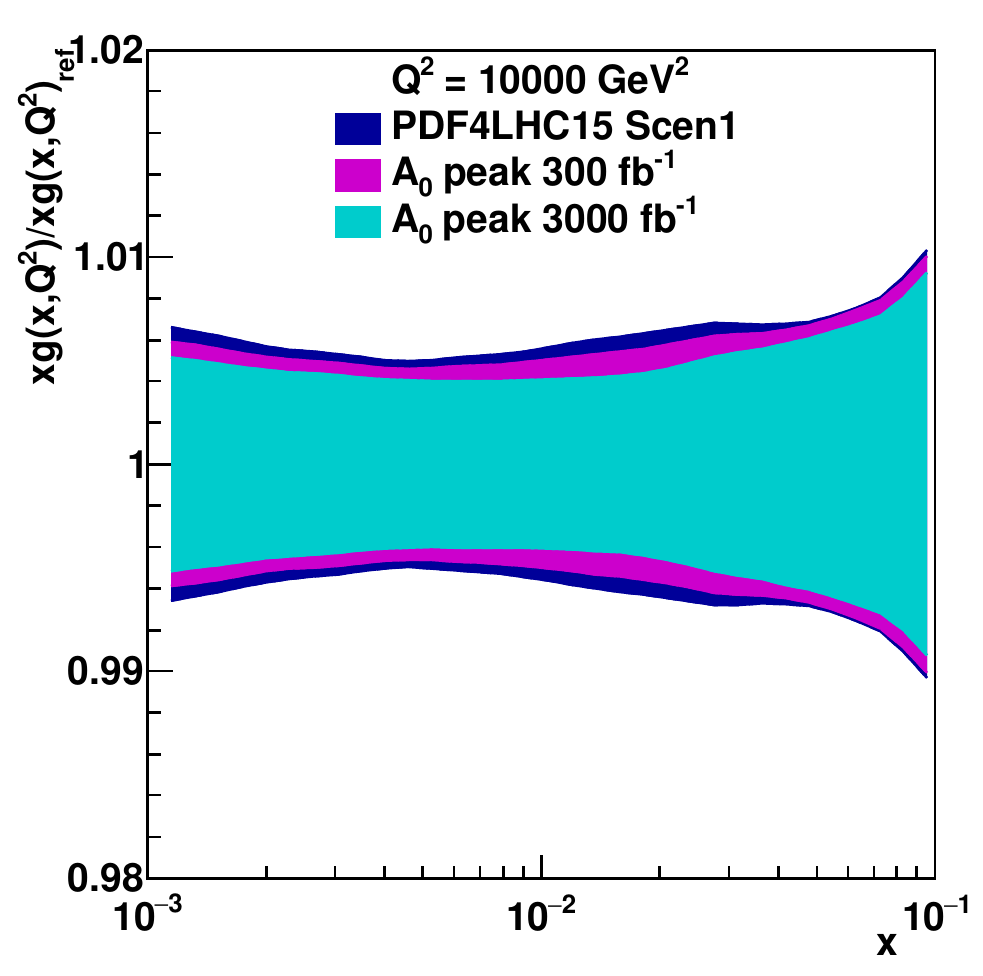}
\end{center}
\caption{Left: The gluon-gluon fusion Higgs boson production cross-section at N$^3$LO for different PDFs, showing the uncertainty from PDFs and their expected reduction including constraints from the $A_0$ measurement assuming 300~fb$^{-1}$ and 3~ab$^{-1}$, respectively. Right: Original PDF4LHC15 Scenario 1 (dark blue) and profiled distributions using $A_0$ pseudo-data corresponding to integrated luminosities of 300~fb$^{-1}$ (magenta) and 3~ab$^{-1}$ (cyan) for 80 $< M <$ 100~GeV and $|Y|<3.5$. Result for gluon ($xg$) is shown for $Q^{2}$ = 10$^{4}$~GeV$^{2}$. Bands represent PDF uncertainties, shown at the 68\% CL.}
\label{fig:HiggsSigma}
\end{figure}

We compute Higgs boson cross section via the gluon-fusion production mechanism for $\sqrt{s}=13$ TeV using the \texttt{MCFM} code~\cite{Campbell:2010ff,Campbell:2019dru} at NLO in QCD perturbation theory and we evaluate PDF uncertainties on this quantity including constraints from $A_0$ profiling. Fig.~\ref{fig:HiggsRapidity} (left) gives the results versus the Higgs boson rapidity $y_H$, where it is visible that the uncertainty is reduced by a factor 2 in the $ - 2 \lesssim y_H \lesssim  2 $ region for the HL-LHC scenario.\\
Furthermore, we compute the gluon-gluon luminosity as a function of invariant mass to assess the reduction of PDF uncertainties. The PDF uncertainties in the gluon-gluon luminosity evaluated at $\sqrt{s}=$13~TeV and computed with CT18nnlo, as well as including constraints from $A_0$ profiling, are depicted in Fig.~\ref{fig:HiggsRapidity} (right). PDF uncertainties are halved in the range 100 $< M_{X} <$ 200~GeV in the HL-LHC scenario.\\ 
Moreover, we evaluate the N$^3$LO Higgs boson total cross section 
using the code \texttt{ggHiggs}~\cite{Bonvini:2014jma,Bonvini:2018ixe}. This cross section and its uncertainty in the cases of the current CT18nnlo~\cite{Hou:2019efy},  NNPDF3.1nnlo~\cite{Ball:2017nwa} and MSHT20nnlo~\cite{Bailey:2020ooq} global sets as well as projected sets, based on complete LHC data sample~\cite{Khalek:2018mdn} is reported in Fig.~\ref{fig:HiggsSigma} (left). Paralleling the profiling analysis of Fig.~\ref{fig:CT18_13TeV_Second}, in Fig.~\ref{fig:HiggsSigma} (right) the profiling analysis which uses projected PDFs based on complete LHC data sample as input~\cite{Khalek:2018mdn} is reported. Remarkably, a reduction in the gluon PDF uncertainty is still visible. In this case, the 300 fb$^{-1}$ to 3 ab$^{-1}$ gain relative to the 
current-to-300 fb$^{-1}$ gain is more significant than in the case of the CT18nnlo~\cite{Hou:2019efy} profiling scenario of Fig.~\ref{fig:CT18_13TeV_Second}. 

\section{Conclusion}
We investigated the implications of precise measurements of  the  angular coefficient $A_0$ near the $Z$-boson mass scale on the theoretical predictions for the Higgs boson production cross section. The same approach, extended to mass regions away from the $Z$ peak, allows the region of larger $x$ momentum fractions to be accessed and will be relevant for associated Higgs boson production with a gauge/Higgs boson or heavy-flavour quarks. Conversely, the extension to low masses can further provide a handle on the small-$x$ regime~\cite{Bonvini:2018ixe,Hautmann:2002tu} of Higgs boson production relevant to the highest energy frontier. Further aspects of the connections between the gauge and Higgs sectors of the SM may be investigated via generalization to the full structure of lepton angular distributions. These other coefficients are generally smaller than $A_0$ and with a milder $p_T$ dependence, but provide a more pronounced $Y$ rapidity dependence.

\end{document}